\documentstyle{l-aa}
\input epsf

\begin{document}
\thesaurus{08    (09.19.2  
                  09.09.1  
                  02.08.1  
                  13.25.4  
                 )
}
\title{Evolution of supernova remnants in the interstellar medium 
       with a large-scale density gradient}
\subtitle{II. The 2-D modelling of the evolution and X-ray emission of  
          supernova remnant RCW86}
\author{O. Petruk}
\institute{Institute for Applied Problems in Mechanics and Mathematics, 
           NAS of Ukraine, 3-b Naukova St., Lviv, 290601, Ukraine\\
	   {Petruk@lms.lviv.ua}
           }
\date{Received ...; accepted ...}
\maketitle


\begin{abstract}
The results of the 2D modelling of Supernova remnant (SNR) 
RCW86 are given. 
Models of this remnant, which for the first time interpret the anisotropy of 
surface brightness as a result of the evolution of adiabatic SNR in the 
interstellar 
medium with a large-scale gradient of density, 
are considered. Estimations on 
the basic characteristics of RCW86 and the surrounding medium 
which follow    
from these models are found. It is shown that the observed surface 
brightness distribution of RCW86 may be obtaned in the 
both proposed up till now models with different initial 
assumptions: one about a Supernova explosion in 185 A.D. and 
another about an explosion in the 
OB-asossiation. In order to obtain the observational contrast 
of surface 
brightness it is necessary to have a medium with the characteristic scale of 
nonuniformity $11\ \mbox{pc}$ if the age of RCW86 is 1800 years or  
$20-25\ \mbox{pc}$ when the SNR is distant from us on $2.8\ \mbox{kpc}$.
The preshock density contrast between the southwestern and northeastern 
parts of RCW86 is in range $3.5-4.5.$ 

\end{abstract}


\medskip

\section{Introduction}

Supernova remnant (SNR) G315.4-2.3 (RCW86, MSH14-6{\it 3}) 
is the result of a supernova explosion which is regarded as 
one of the most possible 
candidates for the "guest star" of 185 A.D. according to Chinese manuscripts 
(Clark \& Stephenson \cite{Clark-Stephenson}). \par

Chin \& Huang (\cite{Chin-Huang}), Schaefer (\cite{Schaefer}) argue 
that the "guest star" in 185 A.D. was not a Supernova (SN). 
Then it is not right 
to assume that the age of RCW86 is $t=1800$ years although the SNR might be 
a relatively young remnant because it has a radial oriented magnetic field 
(Milne \cite{Milne87}) like Tycho, Cas A and Kepler SNRs 
(Strom \cite{Strom}).\par

The SN progenitor might be a member of OB-assosiation. 
In this case we may expect the SN to have been of type II, 
but it could not be born in  
185 A.D. since the SNR has much bigger size than the expected 
one if it is situated at the 
distance 
$d\approx2.8\ \mbox{kpc}$ obtained from the kinematic survey of ionized H 
(Rosado et al. \cite{Rosado-et_al-96}). \par

From the ratio $H_\alpha/H_\beta$ for optical filaments 
Ruiz (\cite{Ruiz}) estimates 
the distance to the SNR as $d\approx 1\ \mbox{kpc}$. \par

RCW86, as radio source MSH14-6{\it 3}, was identified by Hill (\cite{Hill}).  
$\Sigma-D$ relation yields an estimation on the distance to the SNR 
$d=2\div 3.2\ \mbox{kpc}$ (Milne \cite{Milne70}), 
but the possible range is, nevertheless, $d=1\div 10\ \mbox{kpc}$
(Strom \cite{Strom}).   \par

Both the radio and the X-ray observations reveal that the shape of RCW86 
is close to a spherical one with the size $43\arcmin\times 40\arcmin$ (at 
frequency 843 MHz) and is of a shell-like type with the approximately 
axis-symmetrical surface brightness distribution and higher emission 
in the southwest (SW) part of the SNR (Caswell et al.  
\cite{Caswell-Clark}, Pisarski et al. \cite{Pisarski_et_al}, Claas et al.  
\cite{Claas-Smith}, Whiteoak \& Green \cite{Whiteoak-Green}).  
The brightest optical filaments coincide with the maximums of emission in 
radio and X-ray bands (van den Bergh \cite{Bergh}, Pisarski et al.  
\cite{Pisarski_et_al}).  \par

The important results in investigation of RCW86 were caused by analyzing 
the spectral properties of 
X-ray emission from this SNR. The first soft X-ray observations 
were reported  
and interpreted by Naranan (\cite{Naranan}).  Winkler (\cite{Winkler}), 
having used hard 
X-ray data, showed that the observed spectrum might be explained as 
thermal, with two bremsstrahlung components (from the forward and reverse 
shock waves) with the temperatures 
$T_{\rm low}\approx 0.22\ \mbox{keV}$ and $T_{\rm high}\gid 5\ \mbox{keV}.$ \par

The first map of the X-ray surface brightness distribution of RCW86 was presented 
by Pisarski et al. (\cite{Pisarski_et_al}) (EINSTEIN observatory). 
The authors noted that higher emission 
from the SW part might be caused by the 
interaction of the shock wave with the interstellar 
medium (ISM) with a density 2-3 times higher than the average. 
The one temperature 
fitting of the spectrum from the whole remnant (it was accepted that the 
plasma is 
uniform and in collision ionization equilibrium (CIE)) yielded 
temperature $T=1.2\ \mbox{keV}$ and column density 
$N_{\rm H}=2.8\cdot 10^{20}\ {\rm cm^{-2}}$. 
The obtained parameters of RCW86 are given in 
Table~\ref{tab-I}. Pisarski et al. (\cite{Pisarski_et_al}) 
also showed that if this SNR 
is at the adiabatic stage of its evolution, 
it cannot be the result of the type II SN. \par

Nugent et al. (\cite{Nugent_et_al}) analysed the spectrum of RCW86 
(HEAO~1 experiment) with the  
assumptions: 1) that plasma emits in the CIE and, 
2) for the first time, that the plasma 
is under nonequilibrium ionisation condition (NEI) with $T_e=T_i$. The best 
fitting of the CIE spectrum is for 
$T_{\rm high}=5.1\pm 0.14\ \mbox{keV},$ $T_{\rm low}=0.52\pm 0.04\ \mbox{keV}$, $N_{\rm H}=(1.1\pm 
0.3)\cdot 10^{21}\ {\rm cm^{-2}}.$ 
The NEI model gives $N_{\rm H}=(4.4\pm 0.3)\cdot 10^{21}\ {\rm cm^{-2}}$ and 
the parameters of 
the SNR shown in Table~\ref{tab-I}. The distance to RCW86 is smaller than to 
the OB-assosiation but the authors have noted that the 
possible deviation from the NEI 
model may increase the distance to $d=2.5\ \mbox{kpc}$. \par

Claas et al. (\cite{Claas-Smith}) also used both the CIE and the NEI 
model for the 
interpretation of EXOSAT observation of RCW86. 
The possible contamination by the galactic ridge which increases the 
flux above 6 \mbox{keV} was subtracted, 
so in the two-component spectral model 
the smaller 
$T_{\rm high}=3.4\pm 0.2\ \mbox{keV}$ was obtained. 
The low temperature component is believed 
to be a NEI effect. In this paper the parameters of RCW86 were also estimated 
using the 
surface brightness distribution. Higher emission from the SW part must be 
caused by a more dense ISM. Therefore, the Sedov (\cite{Sedov}) solution is 
used only for the interpretation of the northeastern (NE) part of the SNR. 
The NEI model 
yields $T_{\rm s}=T_e/1.3=T_{\rm high}/1.3=(3.03\pm 0.2)\cdot 10^7\ \mbox{K}$. 
\par

ASCA X-ray data of RCW86 are considered by Vink et al. 
(\cite{Vink-Kaastra-B-97}). The authors reveal a remarkable temperature  
variation over the SNR: $T_e=0.8\ \div5\ \mbox{keV}.$ They estimate the 
preshock density contrast for the SW and NE parts of RCW86 as 
$5\div 50$ times.  \par

Modelling the SNR emission it is necessary to take into consideration the 
possibility that the plasma may be under the NEI conditions, 
since the effects of the NEI 
are important. At the same time the nonuniformity of the surrounding ISM is 
very important, too. Calculations show that 
the influence of a nonuniform 
medium on the evolution and the properties of the emission of adiabatic SNRs 
may exceed the influence of the NEI in $10^{1}\div 10^{5}$ times 
(Hnatyk \& Petruk \cite{Hn-Pet97b}, hereafter Paper I). 
At present it is practically inpossible to construct 
models which take into account both the  NEI plasma emission effects and the  
ISM density gradient. Therefore, previous studies of RCW86 are based on the 
one-dimentional Sedov (\cite{Sedov}) solution which cannot restore the 
observed morphology of the SNR. In this paper we 
show that the global anisotropy of the surface brightness distribution 
of RCW86 may be explain as a consequense 
of the SNR evolution in the ISM with a large-scale density gradient. 
We do not involve into the analysis 
the small-scale emission structures which are seen on the SNR's maps, 
because these structures are responsible for the local rises 
and do not essentially modify the global behaviour of the surface brightness. 
The plasma emission model used here is CIE. \par

{
\begin{table*}
\caption[]{Parameters of RCW86 derived from the X-ray observations. 
$R$ is radius of the SNR, $E_{51}$ is the energy of SN explosion in 
$10^{51}\ \mbox{erg},$ $n_{\rm H}^o$ is hydrogen number density of 
surrounding medium, $M$ is swepted up mass, ${\rm M_\odot}$ is the 
mass of Sun. 
          }
\begin{flushleft}
\begin{tabular}{lllllllll}
\hline
Hydrodynamics model$^a$&${\rm PEM^b}$&$t,\ \mbox{years}$& $R,\ \mbox{pc}$ & $d,\ \mbox{kpc}$ & $E_{51}$ & $n_{\rm H}^o,\ {\rm cm^{-3}} $ & $M,\ {\rm M_\odot}$ &  ${\rm Ref.^d}$ \\
\hline
Free expansion (SN~II)                   & {\it CIE}&1800&$9$ &$1.4$&$0.1\div 0.2$&$0.13\div 0.25$&$7\div 14$& [1]\\
Sedov (SN~I)   & {\it CIE}&1800&$6.5$ &$1.4$&$0.1\div 0.2$ &$0.3$&$5$& [1]\\
Free expansion or Sedov & {\it NEI}&$400\div 1900$&$2.5\div9$&$0.4\div 1.6$ &$0.1\div 0.2$&$0.04\div 0.2$&$0.1\div21$&[2]   \\
Sedov for NE part of the SNR & {\it NEI} &1800&6.7&$1.1$ &$0.15$&$0.11^{\rm c}$&$4.9$& [3] \\
\hline
\end{tabular}
\begin{list}{}{}
\item[$^{\rm a}$] All model assumed uniform ISM (density of ISM $\rho^o={\rm const}$) 
\item[$^{\rm b}$] Plasma emission model
\item[$^{\rm c}$] The hydrogen number density of the ISM for the SW part is 
			estimated as 
			$(n^o_{\rm H})_{\rm SW}\sim10(n^o_{\rm H})_{\rm NE}$
\item[$^{\rm d}$] Reference: 
		   [1] -- Pisarski et al. (\cite{Pisarski_et_al}); 
       		   [2] -- Nugent et al. (\cite{Nugent_et_al}); 
         	   [3] -- Claas et al. (\cite{Claas-Smith}).

\end{list}
\end{flushleft}
\label{tab-I}
\end{table*}
}

\section{The 2-D Modelling of RCW86}

\subsection{Method and models}

Evolution of the SNR in the medium with a large-scale density distribution 
is taken as the principal basis for the models presented in this paper.  
We accept here spherically symmetrical energy emission 
during a SN explosion. We assume that the nonspherical shape and the global 
regular surface brightness 
anisotropy of the SNR is caused only by the {\it large-scale} 
stucture of nonuniform medium. \par

The new model parameters associated with 
orientation of the SNR and his projection on the plan of the sky appear 
when 3D modelling executes. 
Projection effects complicate the analysis of the observations.  
In general case, it is impossible to reproduce unambiguily the real 3D shape 
for the knowing projection. In case of considered here the axis-symmetrical 
models, projection effects can easily be included into model. Simple 
connections between the parameters of the model and the projection exists. 
The angle $\delta$ between the symmetry axis of the 2D SNR and the visual 
cross-section is a new additional free paremeter of the model. \par

At our previous paper ({\it Paper I}) 
it have been described a new approximate analytical method which 
allows the modelling a point explosion in media with a large-scale density 
gradient.  
We apply this method here to the discribing of an evolution of the SNR RCW86. 
The possibilities of the SN 
explosion in 185 A.D. and in the OB-assosiation will be considered 
separately . \par

We take as the basic initial model's parameters the three observational 
results: the visual angular size of the SNR $\Theta\approx 40\arcmin$, 
the temperature of the 
plasma's hot component $T_{\rm high}=3.4\pm 0.2\ \mbox{keV},$ which 
corespond to corrected for the contribution of the galactic ridge 
emission spectrum from the whole SNR (Claas et al.  \cite{Claas-Smith}) and  
the surface brightness distribution profile along the symmetry axis of 
the SNR's image  (Pisarski et al. \cite{Pisarski_et_al}) 
(see Fig.~\ref{rcw6}, \ref{rcw1}). This profile is sensitive to 
large-scale density distribution of ISM and allows to see the properties of 
the density gradient. \par

These characteristics are supplemented additionally with the fourth  
one depending on the one from two basic assumptions:  
the first assumption is that RCW86 is a result of the SN explosion in 185 A.D. 
(that gives the age of the SNR) and 
the second one is that RCW86 is a result of the SN explosion in the 
OB-assosiation (this yields the distance to the SNR). \par 

The temperature $T_{\rm s}$ at the shock front in the 
Sedov (\cite{Sedov}) model of the SNR may be 
obtaned from $T_{\rm high}$: 
$T_{\rm s}\approx T_{\rm high}/1.3$ (Itoh \cite{Itoh}).  In 
Paper I it have been shown that the nonsperical SNR may be characterised 
by the some "average" characteristic temperature $T_{\rm ch}\propto 
\left(M\right)^{-1},$  
which will be likely connected with $T_{\rm high}$: 
\begin{equation}   
T_{\rm ch}\approx T_{\rm high}/1.3=(3.03\pm 0.2)\cdot 10^7\ \mbox{K}. 
\end{equation} 

We have also shown that during the adiabatic stage the value of  
$T_{\rm ch}$ is 
close to $T_{\rm s}$ of the Sedov SNR with the same initial model 
parameters (e.g., the 
energy of explosion $E_{o},$ the hydrogen number density 
in the place of explosion $n^o_{\rm H}(0)$) 
with the maximal error about $20\div 30\%$. Therefore, 
we may write analoguously to the Sedov case that 
\begin{equation}
\label{Tch_Ts_t}
T_{\rm ch}\approx T_{\rm s}=2.08\cdot 10^{11} \left({E_{51} \over n_{\rm H}^o(0)
}\right)^{2/5} t_{yr}^{-6/5}\ \mbox{K}, 
\end{equation}
\begin{equation}
\label{Tch_Ts_R}
T_{\rm ch}\approx 6.47\cdot 10^{9} \left({E_{51} \over n_{\rm H}^o(0)
}\right) (\overline{R}_{\rm pc})^{-3}\ \mbox{K}, 
\end{equation}
where $E_{51}=E_o\cdot10^{-51},$ $t_{yr}$ is the age of the SNR in years, 
$\overline{R}_{\rm pc}$ is the average radius of the nonspherical SNR 
in \mbox{pc}. Hereafter $\gamma=5/3$.  

For the calculation of the X-ray emission of the plasma in the CIE 
the Raymond \& Smith (\cite{Raym-Smith77}) data were approximated. \par

\subsection{RCW86 as the result of the SN explosion in 185 A.D.}
\subsubsection{Exponential medium}

Let us take as initial SNR characteristics the angular size $\Theta,$ 
the temperature $T_{\rm ch},$ the surface brightness profile and 
additionally the age of the SNR $t=1800\ \mbox{years}.$ \par

We will consider the evolution of the SNR in the nonuniform medium with 
the flat exponential density distribution 
\begin{equation}     
\label{exp-density}
\rho^o(r,\theta)=\rho^o(0) \exp\left(-{r\over H}\cos\theta\right),
\end{equation} 
where $\rho^o(0)$ is the initial density around the place of SN explosion, 
$H$ is the scale of the density nonunifirmity, $\theta$ is 
an angle between the maximal density decreasing direction 
and the considered direction. 
Such exponential distribution is the one of the most possible to approximate 
real continuous density distributions in interstellar clouds, gaseous 
galactic disc etc. \par

\begin{figure}[t]
\epsfysize=9.2truecm
\epsfbox{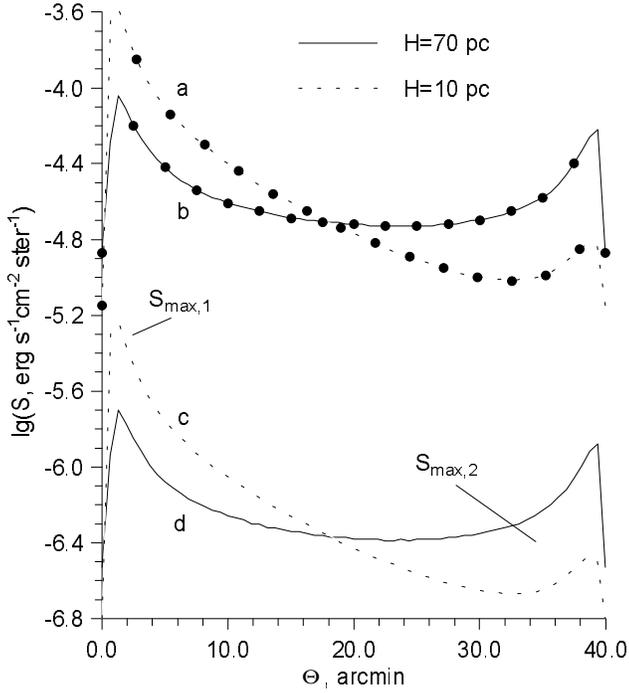}
\caption[]{Surface brightness $S$ distribution in the diapason 
$\varepsilon=0.1\div2\ \mbox{keV}$ along of the median of the SNR 
in the ISM with the exponential density distribution (\ref{exp-density}) 
for the four cases of the initial model parameters. 
It is suppose here for all cases that ${E_{51}/n_{\rm H}^o(0)}=1.5,$ 
$H_{\rm pc}\tau^{2/5}=6.85,$ $t=1800\ \mbox{years}$ and 
$T_{\rm ch}=3.0\cdot 10^7\ \mbox{K}.$ 
Solid lines show the cases with $H=70\ \mbox{pc},$ 
dashed -- the cases with $H=10\ \mbox{pc}.$ Energies of explosion: 
a, b: $E_{51}=1;$ c, d: $E_{51}=0.15.$  
Dots represent the cases c, d multiplied by the factor 
$((n^o_{\rm H}(0))_{top}/(n^o_{\rm H}(0))_{bottom})^2=10^{1.66}.$ 
          }
\label{rcw4}
\end{figure}
 
\begin{figure}[t]
\epsfysize=6.9truecm
\epsfbox{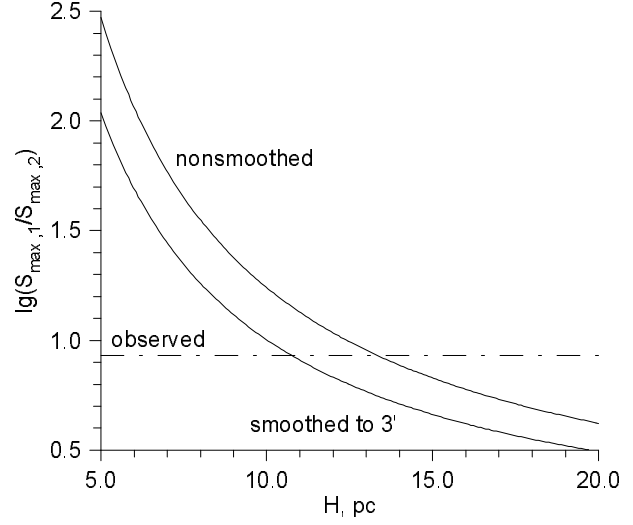}
\caption[]{Ratios of the values of the two peaks in the surface 
brightness $S$ distribution versus the high scale $H$ for models of SNR 
in the flat exponential media (\ref{exp-density}).
Here $t=1800\ \mbox{years}$ and ${E_{51}/n_{\rm H}^o(0)}=1.5.$ 
In the figure both contrast of surface brightness are shown: smoothed to 
$3\arcmin$ like observed (Pisarski et al. \cite{Pisarski_et_al}) and without 
smoothing.
          }
\label{rcw5}
\end{figure}

\begin{figure}
\epsfysize=7.1truecm
\epsfbox{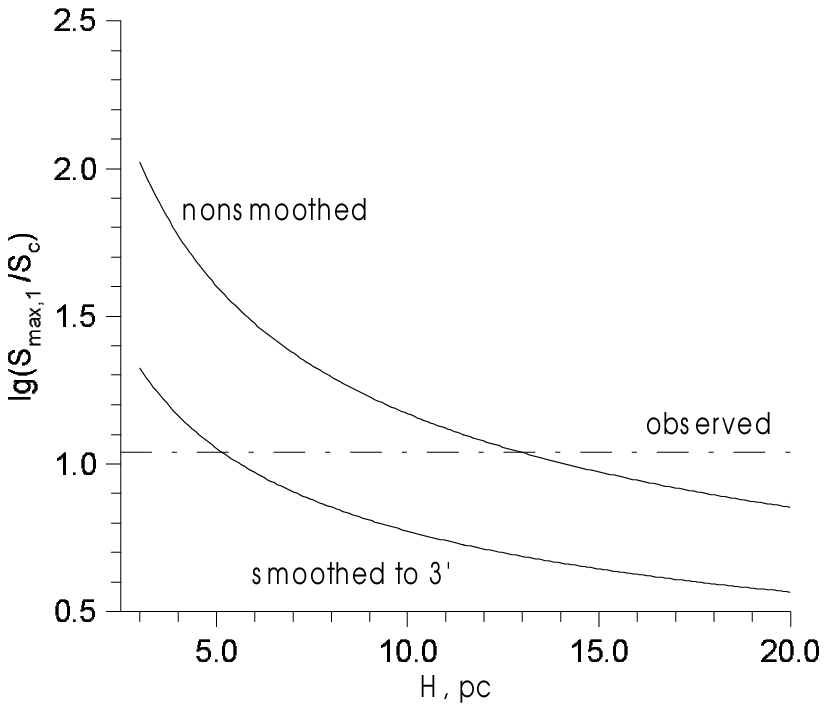}
\caption[]{The same as in Fig.~\ref{rcw5} for the ratios 
between the maximum in the surface brightness distribution $S_{max,1}$ 
and the value of the brightness in the visual geometric center of the 
SNR $S_{c}$.
          }
\label{rcw5b}
\end{figure}

When we fixed $T_{\rm ch}$ and $t$ we 
have got from (\ref{Tch_Ts_t}) an estimation on 
\begin{equation}
\label{estim-E_n-2}
{E_{51}/n_{\rm H}^o(0)}=1.5\pm 0.25.
\end{equation}
From this relation and appropriate explosion energy range 
$E_{51}=0.1\div 1$ the range for the initial number density 
$n_{\rm H}^o(0)=0.06\div 0.8\ {\rm cm^{-3}}$ follow. \par

From the definition of a dimentionless time 
(Hnatyk \& Petruk \cite{Hn-Pet96}) $\tau=t/t_{\rm m}$ with 
$t_{\rm m}=\alpha_{\rm A}^{1/2}E_o^{-1/2}\rho_o^{1/2}(0)R_{\rm m}^{5/2}$ 
(where $\alpha_{\rm A}$ is the self-similar constant, $R_{\rm m}$ is the distance scale) 
it is obtaned for $\gamma=5/3$ that   
\begin{equation}
\label{t_t_m}
t=\tau t_{\rm m}=18.01\left({E_{51} \over n_{\rm H}^o(0)}\right)^{-1/2}
{R_{{\rm m},{\rm pc}}}^{5/2}\tau\quad \mbox{yr},
\end{equation}
$R_{{\rm m},{\rm pc}}$ is the distanse scale in pc. \par

From this relation and (\ref{estim-E_n-2}) a connection between 
the scale high $H$ and the dimentionless time follows:  
\begin{equation}
\label{conn_H_tau}
H=R_{\rm m}\simeq (6.8\pm 0.2)\cdot \tau^{-2/5}\quad \mbox{pc}. 
\end{equation}

The possible values of $H$ lie in the range $H=150\ \mbox{pc}$ for 
the Galactic's disk to few parsecs for a local nonuniformity of 
the ISM. 
The influence of the nonuniform medium on the dynamics of the SNR become more 
apprecieble 
with the age of a SNR. The surface brightness distribution of a SNR 
will be close to the Sedov one for small $\tau$. 
Since the observational distribution of the surface brightness in RCW86 
is not similar to the Sedov one, 
the dimentionless time $\tau$ have not to be small. \par

The observed ratio $\xi$ between the axes of the visual shape of RCW86 
is in range $\xi=1.0\div 1.2$ (Pisarski et al. \cite{Pisarski_et_al}). 
This corresponds to ranges of $\tau=0.01\div 11$ 
(Fig.~5 in Paper I). The scale hight $H$, as it follow from 
(\ref{conn_H_tau}), then must be anywhere between 2.6 and 40 pc. Such wide 
range of $H$ is a result of very slow decreasing of the visual shape's 
anizotropy with time (the same Fig.). 
Small scale hight (like $H=2.6\ {\rm pc}$) it seems to be impossible because 
the surface brightness contrast then must be about $10^{4.7}$ times. 
Possible presence in the shape anizotropy the residual component related 
to anizotropy which the SNR might have on the free expansion stage 
(due to ashperical explosion) is the second reason which complicates 
using the small anizotropy of the shape for estimating the $H$ and $\tau.$ \par

Physically the most correct limitation on $H$ and $\tau$ we may get from 
the profiles of the surface brightness. 
In Fig.~\ref{rcw4} are shown the profiles of the surface brightness for the 
four 
cases of the initial model peremeters. Relations (\ref{estim-E_n-2}) and 
(\ref{conn_H_tau}) have place for each set of the parameters. We may see that 
for fixed $H$ and $\tau$ the shape of the profile is the same. Value of the 
blast 
energy $E_{51}$ (and as consequence of (\ref{estim-E_n-2}), the number density 
$n_{\rm H}^o(0)$) give the amplitude of the profile which may differ with the 
factor about 
few ten times. It is follow from Fig.~\ref{rcw4} that the values $H$ 
and $\tau$ affect on the contrast of the surface brightness. \par

For finding $H,$ it is possible to accept the contrast between the brightness 
in the different points of the SNR's image. For example, it may be used 
the relation 
between the two "near-front" maximums in the distribution. 
The most appropriate value of high scale for this choice, 
as it may be seen from Fig.~\ref{rcw5}, 
is $H=11.0\ \mbox{pc}$ 
(and respectively $\tau=0.3$). Then calculations show (Fig.~\ref{rcw6}) that 
the most appropriate value of the blast energy is $E_{51}=0.22$ 
($n_{\rm H}^o(0)=0.15\ {\rm cm^{-3}}$). \par

For normalizing the surface 
brightness $S$ in photon energy range $\varepsilon=0.1\div2.0\ \mbox{keV},$  
the effective values of photon energy $\overline{\varepsilon}=0.1\ \mbox{keV}$ 
and absorption cross section $\overline\sigma$ as cross section at photon 
energy ${\varepsilon}=0.316\ \mbox{keV}$ from  
Morrison \& McCammon (\cite{Morr-McCammon83}) have been used. \par

Obtaned SNR's luminocity $L_{\rm x}^{0.1-2}=3.5\cdot 10^{34}\ {\rm erg/s}$ 
is consistent with esteblished by Pisarski et al. (\cite{Pisarski_et_al}) 
$L_{\rm x}^{0.2-4}=2\cdot 10^{34}d_{\rm kpc}^2\ {\rm erg/s}.$ \par 

In Fig.~\ref{rcw3new} it is shown the maps of the surface brightness 
distribution for this model in comparison with the distribution of the 
spectral index 
$\alpha=-\partial \ln F_\varepsilon/\partial \ln \varepsilon,$ where  
$F_\varepsilon$ is the continuum flux at photon energy $\varepsilon$. \par 

\subsubsection{Power law medium}

It is useful to note, that concreete appearance of surrounding density 
distribution is not strongly essential then size of SNR does not exceed 
a few scale hights. Namely, 
similar parameters of model may be obtained 
under assumption that the SNR evolve in other type of media, if 
contrast of density along the surface of the SNR will be similar. 
Let us consider, for example, a medium 
with the spherically-symmetrical power law density distribution created 
by steller wind 
\begin{equation}
\label{r^w}
\rho^o(\tilde r)=\rho_o(\tilde r/R_{\rm m})^{\omega},
\end{equation}
then the SN explosion position is displaced on the 
distance $r_o$ from the center of symmetry $\tilde r=0$, 
therefore, the density 
distribution as a function of the distance $r$ from the 
explosion point $r=0$ is:  
\begin{equation}     
\label{r^w-density}
\rho^o(r,\theta)=\rho^o(0){\left( 
{\sqrt{r_o^2+r^2+2rr_o\cos\theta}\over r_o}    
\right)}^{\omega}.                                                    
\end{equation} 

We take $\omega=-2$ and calculate a number of models in 
order to get an appropriate $r_o$ using the surface brightness contrast 
between the two maximums in the brightness distribution. \par

\begin{figure}
\epsfysize=7.1truecm
\epsfbox{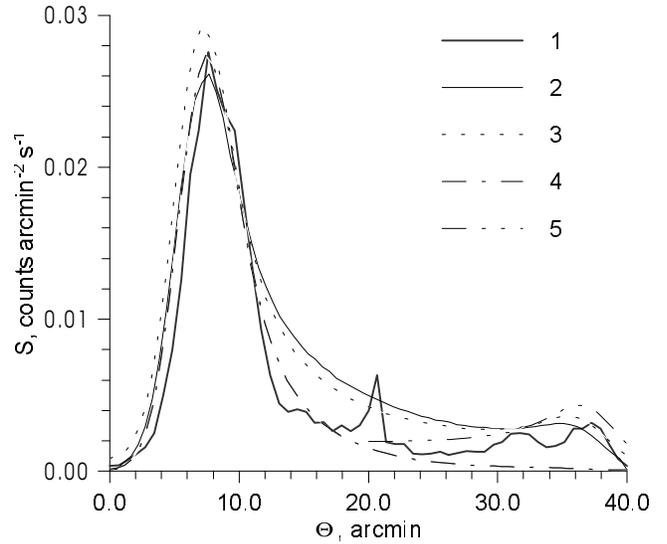}
\caption[]{Observed (line 1) distribution of the 
surface brightness $S$ along of the 
symmetry axis of RCW86 (Pisarski et al. \cite{Pisarski_et_al}) and the 
distributions in the models of SNRs with the age 
$t=1800\ \mbox{years}$: line 2 --  
the exponential medium (\ref{exp-density}) with $H=11\ \mbox{pc}$ and 
$E_{51}=0.22,$ $n^o_{\rm H}(0)=0.15\ {\rm cm^{-3}};$ 
line 3 -- the power law medium (\ref{r^w-density}) with $r_o=21.8\ \mbox{pc}$ 
and the same $E_{51},$ $n^o_{\rm H}(0);$ 
line 4 -- the exponential medium (\ref{exp-density}) with $H=5\ \mbox{pc}$ 
and $E_{51}=0.17,$ $n^o_{\rm H}(0)=0.11\ {\rm cm^{-3}};$ 
line 5 --  the uniform medium, $E_{51}=0.17,$ $n^o_{\rm H}=0.11\ {\rm cm^{-3}}.$ 
All distributions are smoothed to $3\arcmin.$
          }
\label{rcw6}
\end{figure}

\begin{figure}
\epsfysize=13.53truecm
\epsfbox{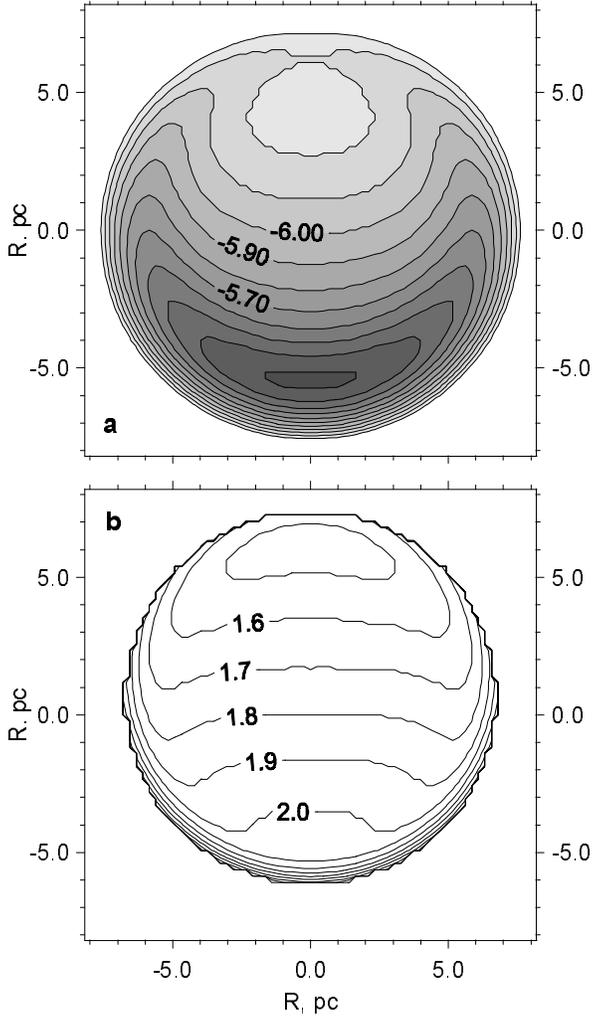}
\caption[]{{\bf a} The surface brightness $S$ 
(in ${\rm erg\ s^{-1}\ cm^{-2}\ st^{-1}}$) 
distribution in the photon energy range $\varepsilon=0.1-2\ \mbox{keV}.$ 
{\bf b} The surface distribution of the spectral index $\alpha$ at 
$\varepsilon=5\ \mbox{keV}$ for the SNR model in 
the exponential medium (\ref{exp-density}) with 
$H=11.0\ \mbox{pc}.$ Parameters of the model are 
$t=1800\ \mbox{years},$ $E_{51}=0.22,$ $n^o_{\rm H}(0)=0.15\ {\rm cm^{-3}}$ 
and $\delta=0^o.$ 
          }
\label{rcw3new}
\end{figure}

Observed surface brightness contrast is obtaned in this model with 
$r_o=21.8\ {\rm pc}.$ 
Characteristic scale $H_{\rm ch}$ for medium (\ref{r^w-density}) with such 
$r_o$ may be estimated. It equals $H_{\rm ch}=11\ {\rm pc}.$ 
Other parameters of the model are really close to the same parameters of the 
SNR in exponential medium (\ref{exp-density}) and shown in 
Tables~\ref{tab-ii} and Fig.~\ref{rcw6}. \par

\subsubsection{Exponential plus uniform medium}

RCW86 may evolve in the ISM with the contact between the uniform 
medium 
$\rho^o={\rm const}$ and the region with the higher density 
(e.g., interstellar cloud) 
where density is distributed according to the exponential law 
(\ref{exp-density}). In this case one part of the SNR (the NE part of RCW86) 
will evolve in the uniform medium whereas the evolution of the another part 
(the SW part of RCW86) will be 
determined by the variation of the density in the dense region. 
In this composite model 
the surface 
brightness in the central part of the SNR will be close to the brightness 
which the Sedov 
model gives, but the maximal brightness will be determined by the concrete 
distribution of the density in the nonuniform region. \par

When we take as initial parameter the ratio of the  
maximum of the brightness distribution to the value of the brightness 
in the visual 
geometrical centre of the SNR, we obtaine another model parameters for 
the case of the exponential law density distribution, namely, 
$H=5.0\ \mbox{pc}$ 
(Fig.~\ref{rcw5b}) and $E_{51}=0.17$ (Fig.~\ref{rcw6}). We may calculate now 
the profile of the surface brightness for the Sedov model wich 
discribes the NE part of RCW86 
in the assumption of the composite medium (Fig.~\ref{rcw6}). 
The initial parameters for it are the same as for the SW part, 
$E_{51}=0.17,$ $n^o_{\rm H}(0)=0.11\ {\rm cm^{-3}}.$  We may see that the 
composite medium model (exponential plus constant 
density distributions) more better describes the observational 
surface brightness distribution than the exponential medium alone. \par

\begin{table}
\caption[]{Parameters of RCW86 obtaned from the different models 
of the SNR. Plasma emission model is CIE. $L_{\rm x}$ is in the photon 
energy range $\varepsilon=0.1\div2.0\ \mbox{keV}.$ 
$R(0)$ and $R(\pi)$ are the radii of shock in directions $\theta=0$ and 
$\theta=\pi$ respectively, $D$ is the shock velocity.
          }
\begin{flushleft}
\begin{tabular}{lccccc}
\hline
Parameters&\multicolumn{5}{c}{Initial density distribution}\\
\cline{2-6}
of model                          &${\rm E^a}$&${\rm PL^b}$&${\rm EU^c}$&${\rm E^a}$&${\rm E^a\ }$ \\
\hline
\multicolumn{6}{l}{\it Observed}\\ 
$\Theta,$\ arcmin                 &40         &40  &40  &40           &40\\ 
$T_{\rm ch},$\ $10^{7}$ \mbox{K}  &3.0        &3.0 &3.0 &3.0          &3.0\\ 
\hline
\multicolumn{6}{l}{\it Additionally supposed}\\ 
$t,$\ \mbox{years}                &1800       &1800&1800&---          &--- \\ 
$d,$\ \mbox{kpc}                  &---        &--- &--- &2.8          &2.8\\ 
$H,$\ \mbox{pc}                   &11         &$\sim 11^d$&5.0   &26  &22  \\
$\delta,$\ degree                 &0          &0   &0   &0            &30  \\
\hline
\multicolumn{6}{l}{\it Obtained}\\ 
$t,$\ \mbox{years}                &1800       &1800&1800&4300         &4300         \\
$E_{\rm o},$\ $10^{51}$ \mbox{erg}&0.22       &0.22&0.14&2.0          &2.0          \\
$n^o_{\rm H}(0),$\ ${\rm cm^{-3}}$&0.15     &0.15&0.10&0.1          &0.1        \\
$R(0),$\ \mbox{pc}                &7.50       &7.42&6.84&17.96        &18.25        \\ 
$R(\pi),$\ \mbox{pc}              &6.27       &6.20&5.78&14.96        &14.78        \\ 
$D(0),$\ \mbox{km/s}              &1800       &1740&1490&1810         &1870         \\ 
$D(\pi),$\ \mbox{km/s}            &1260       &1220&1070&1250         &1230         \\ 
$T(0),$\ $10^{7}$ \mbox{K}        &4.45       &4.18&3.03&4.50         &4.82          \\ 
$T(\pi),$\ $10^{7}$ \mbox{K}      &2.17       &2.03&1.58&2.16         &2.07         \\ 
$n^o_{\rm H}(R,0),$\ ${\rm cm^{-3}}$ &0.08    &0.08&0.10&0.18         &0.19         \\
$n^o_{\rm H}(R,\pi),$\ ${\rm cm^{-3}}$ &0.27  &0.29&0.31&0.05         &0.04        \\
$d,$\ \mbox{kpc}                  &1.18       &1.17&1.08&2.83         &2.83         \\
$M,$\ ${\rm M_\odot}$             &6.9        &6.8 &--- &62           &62           \\
$lg(L_{\rm x}, \mbox{erg/s})$     &34.54     &34.53&--- &35.33        &35.35        \\
$\alpha(5\ \mbox{keV})$           &1.90       &1.88&--- &1.91         &1.92         \\
\hline
\end{tabular}
\begin{list}{}{}
\item[$^{\rm a}$] E is the flat exponential medium (\ref{exp-density})
\item[$^{\rm b}$] PL is the power law medium (\ref{r^w-density}) with 
		$\omega=-2$ and $r_o=21.8\ {\rm pc}$
\item[$^{\rm c}$] EU is the exponential medium (\ref{exp-density}) plus 
			uniform 
\item[$^{\rm d}$] Characteristic scale hight $H_{\rm ch}$
\end{list}
\end{flushleft}
\label{tab-ii}
\end{table}

It is nesessary to note, when RCW86 is the SNR created by the 
explosion in 185 A.D., 
we see it near the maximum disclosing projection ($\delta\approx 0^o$). 
Really, another 
projections hide the real contrasts (e.g., of axes size, surface brightness; 
see Paper I). If we see the contrast of surface brightness decreased 
already, then the real one must be greater and therefore $H$ must be 
smaller. \par

Obtaned parameters of RCW86 are summarised in Table~\ref{tab-ii} (the first 
three columns). \par 

\subsection{RCW86 as the result of the SN explosion in the OB-association}

Let us suppose now that RCW86 is a result of the SN explosion in the 
OB-assosiation which distance from us is estimated (e.g., Westerlund 
\cite{Westerlund}). In this section the next question is the main: 
what scale hight has the medium around the RCW86 if the SNR's progenitor 
have exploded in this OB-assosiation? Therefore the nonuniform media with 
exponential density distribution (\ref{exp-density}) is used. \par
 
We take as the initial SNR characteristics the observational angular size 
$\Theta,$ the temperature $T_{\rm ch},$ the surface brightness profile 
and, additionally, the distance to the remnant 
$d\approx 2.8\pm 0.4\ \mbox{kpc}$ (Rosado et al. \cite{Rosado-et_al-96}). \par

\begin{figure}
\epsfysize=7.2truecm
\epsfbox{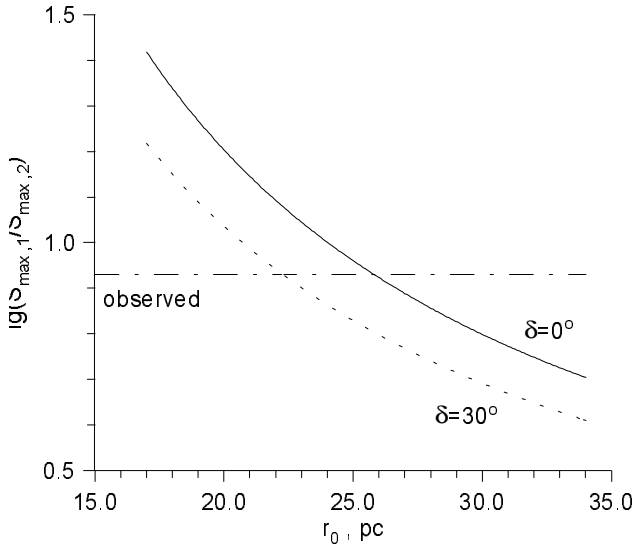}
\caption[]{
Ratios of the values of the two peaks in the X-ray surface 
brightness $S$ distribution versus $H$ for the SNRs in the 
exponential media (\ref{exp-density}), when $t=4300\ \mbox{years},$ 
${E_{51}/n_{\rm H}^o(0)}=20.3$ and $\delta=0^o.$ 
Dashed line is the same relation for the same SNR models, except $\delta=30^o.$
All ratios are smoothed to $3\arcmin$.
          }
\label{rcw1a}
\end{figure}

It is possible now to estimate the average SNR's radius from $\Theta$ and 
$d$: $\overline{R}_{\rm pc}=0.5d_{\rm kpc}\Theta\arcmin/3.438\approx 16.3\pm 2.3\ 
\mbox{pc}.$
Therefore we have got from (\ref{Tch_Ts_R}) an estimation on 
\begin{equation}
\label{estim-E_n-1}
{E_{51}/n_{\rm H}^o(0)}\simeq 23\pm 10.
\end{equation}
This means that the explosion energy 
must be high and the initial number density have to be low. \par

The age of the SNR as it follows from (\ref{Tch_Ts_t}) is anywhere 
from $t=3500\ \mbox{years}$ for $T_{\rm ch}=2.8\cdot10^7\ \mbox{K}$ and 
${E_{51}/n_{\rm H}^o(0)}=13,$
up to $t=5350\ \mbox{years}$ in case $T_{\rm ch}=3.2\cdot10^7\ \mbox{K}$ 
and ${E_{51}/n_{\rm H}^o(0)}=32.$
We may see, with agreement with Rosado et al. (\cite{Rosado-et_al-96}), 
if RCW86 is the remnant of the SN exploded in 
OB-assosiation it can not be borned in 185 A.D. \par

\begin{figure}
\epsfysize=7.2truecm
\epsfbox{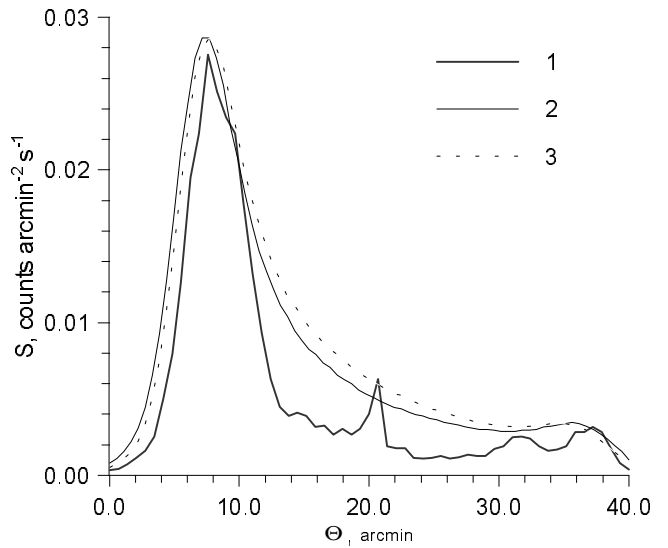}
\caption[]{Distribution of the surface brightness $S$ 
along the symmetry axis of the SNR RCW86: line 1 -- 
observational (Pisarski et al. \cite{Pisarski_et_al}); 
line 2 -- model in exponential medium (\ref{exp-density}) with 
$H=25.7\ \mbox{pc}$ and $t=4300\ \mbox{years},$ $E_{51}=2.0,$ 
$n^o_{\rm H}(0)=0.1\ {\rm cm^{-3}},$ $\delta=0^o;$ 
line 3 -- the same model, except $H=22.3\ \mbox{pc}$ and $\delta=30^o.$ 
All distributions are smoothed to $3\arcmin.$
          }
\label{rcw1}
\end{figure}

For the distance $d=2.8\ \mbox{kpc}$ and $T_{\rm ch}=3.0\cdot10^7\ \mbox{K}$ 
the age $t=4300\ \mbox{years}$ and ratio ${E_{51}/n_{\rm H}^o(0)}=20.3$ 
have been obtained. We calculate the evolution of RCW86 in the exponential 
medium (\ref{exp-density}) with these initial parameters. \par

It may be seen from Fig.~\ref{rcw1a} that the scale hight $H=25.7\ \mbox{pc}$ 
is the most appropriate for the case of the maximal disclosure of the 
remnant ($\delta=0^o$). \par

\begin{figure}
\epsfysize=13.50truecm
\epsfbox{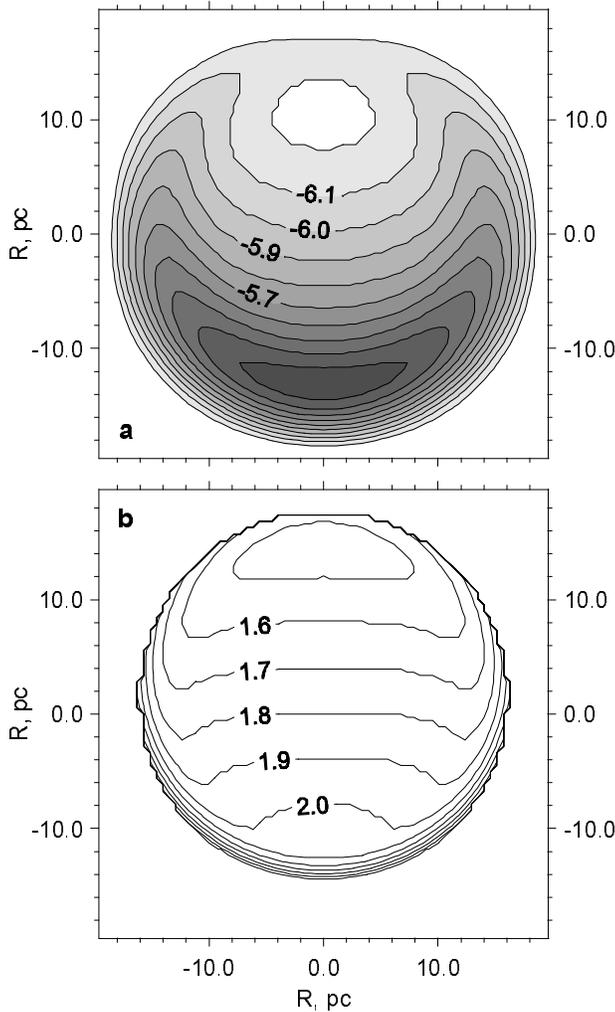}
\caption[]{{\bf a} The surface brightness $S$ 
(in ${\rm erg\ s^{-1}\ cm^{-2}\ st^{-1}}$) distribution in the range 
$0.1-2\ \mbox{keV}$ 
and {\bf b} the surface distribution of the spectral index $\alpha$ at 
$\varepsilon=5\ \mbox{keV}$ for the SNR model in 
the nonuniform exponential medium (\ref{exp-density}) with 
$H=25.7\ \mbox{pc}.$  Parameters of the model are 
$t=4300\ \mbox{years},$ $E_{51}=2.0,$ $n^o_{\rm H}(0)=0.1\ {\rm cm^{-3}}$ 
and $\delta=0^o.$ 
          }
\label{rcw3}
\end{figure}

In order to estimate an explosion energy $E_o$ and a number 
density $n_{\rm H}^o(0)$ it was calculated a number of the models with 
such $H$ and different $E_o.$ It may be seen from 
Fig.~\ref{rcw1} that the surface 
brightness distribution along the symmetry axis for $E_{51}=2.0,$ 
$n_{\rm H}^o(0)=0.1\ {\rm cm^{-3}}$ is near of the observed one. 
The whole maps of the surface brightness $S$ and the spectral index $\alpha$ 
distribution are shown in Fig.~\ref{rcw3}. \par

Fig.~\ref{rcw1} demonstrates also 
the surface brightness distribution along the symmetry axis for 
the case of the different projection of the SNR, 
when $\delta=30^o$. We see that distribution, like observational, may be 
obtaned also under different projection of the 2D SNR on the plan of the sky
with the smaller $H=22.3\ \mbox{pc}.$ Now we cannot separate one case of 
projection from another but we see also that the model with $\delta=30^o$ 
does not essentially change the SNR's parameters which are summarised for 
these different models of RCW86 in the last two columns of 
Table~\ref{tab-ii}. \par

Modelling reveal also that for model with $\delta=30^o$ 
both the surface brightness distribution and the surface distribution 
of the spectral index are also very close to shown in Fig.~\ref{rcw3} 
case of $\delta=0^o$. \par

\subsection{Contribution of the ejecta}

Nugent et al. (\cite{Nugent_et_al}) have made the conclusion that 
the observed data on RCW86 can be fitted without taking into consideration 
the emission from the ejecta. These authors have also suggested that the 
value of $\beta=M_{\rm ej}/E_{51}$ ($M_{\rm ej}$ is the mass of ejecta in 
${\rm M_\odot}$) lies much probably 
in the range 1 to 10. In our models of the SNR the blast energy are 
$E_{51}=0.22$ in the case of SN explosion in A.D.185 and $E_{51}=2.0$ in the 
case of SN explosion at a distance $d=2.8\ {\rm kpc}.$ 
Terefore, the mass of ejecta may be estimated as 
$M_{\rm ej}=0.22\div 2.2{\rm M_\odot}$ or $M_{\rm ej}=2\div 20{\rm M_\odot}.$  
Swepted up masses are respectively 7 and 62 ${\rm M_\odot}$ and  
exceeds $M_{\rm ej}$ in 3 to 31 times. We expect that 
under our assumption of CIE conditions and obtained $n^o_{\rm H}\sim 0.1\ 
{\rm cm^{-3}},$ such swepted up masses are enough for RCW86 to be in Sedov 
phase of his evolution. So, we may assume that in both cases the ejecta do 
not considerably modify the emission from RCW86. 
However, if to take into account the NEI 
effects and consider the value of the $\beta$ up to the possible upper limit 
which have been put as $\beta\sim 40,$ the contribution of the ejecta into 
the emission might be essential (Nugent et al. \cite{Nugent_et_al}). 
On the other hand, if RCW86 would evolve in very dense ISM 
($n^o_{\rm H}\sim 10^3$), it would entry in Sedov stage 
after swepting up $M\sim 19M_{\rm ej}$ (Dohm-Palmer \& Jones 
\cite{Dohm-P_Jones}). \par

\section{Conclusions}

Observations reveal complicated RCW86 morphology. Close to spherical shape 
of the SNR coexists with the surface brightness distribution 
which is far from the Sedov one.  When we take into 
account the nonuniform ISM we may explain such morphology. \par 

We considered here the 2-D models of RCW86 in a few types of 
nonuniform media. It is shown that the features of the surface brightness 
distribution allow us to restore the SNR characteristics. As it is described, 
the contrasts in the surface brightness depend on the gradient of the density 
of the ISM and 
the age of the SNR, thereas the amplitude of the surface brightness 
distribution depends 
on the energy of the Supernova explosion and the initial density 
around the place of explosion. \par

It is shown that observed surface brightness 
distribution of RCW86 may be obtaned in the models with the two 
different initial 
assumptions: one about the Supernova explosion in 185 A.D. and another 
about the 
explosion in the OB-asossiation. Data we posses do not allow us to diside 
which of those models is true. The parameters obtaned for RCW86, 
basing on these two assumption, are summarased in Table~\ref{tab-ii}. 
The models give the observational contrasts of the surface 
brightness when to consider the ISM with the scales of 
nonuniformity $11\ \mbox{pc}$ (if the age of RCW86 is 1800 years) or 
$20-25\ \mbox{pc}$ (when the SNR is distant from us on 2.8 kpc). 
The preshock density contrasts 
between the southwestern and the northeastern parts of RCW86 in range 
$3.5-4.5$ are obtained for both considered assumptions.\par
 

\begin{acknowledgements}
I am very grateful to Dr.~Bohdan Hnatyk for the permanent attention 
to my work as well as for always usfull advices. 
\end{acknowledgements}



\end{document}